\documentclass[fleqn,usenatbib,useAMS]{mnras}
\usepackage{graphicx}	
\usepackage{amsmath}	
\usepackage{amssymb}	
\usepackage{multicol}   
\usepackage{bm}		
\usepackage{pdflscape}	
\usepackage{appendix}
\usepackage{ulem}
\usepackage{tcolorbox}

\usepackage{multibib}
\newcommand{\beq}{\begin{equation}}
\newcommand{\eeq}{\end{equation}}
\newcommand{\bse}{\begin{subequations}}
\newcommand{\ese}{\end{subequations}}
\newcommand{\bary}{\begin{eqnarray}}
\newcommand{\eary}{\end{eqnarray}}
\newcommand{\bwt}{\begin{widetext}}
\newcommand{\ewt}{\end{widetext}}

\usepackage[T1]{fontenc}
\usepackage{ae,aecompl}

\title{Constraining the redshift of BL Lac VER J0521+211}

\author[S. Sahu et al.]{
Sarira Sahu$^{1}$%
\thanks{Contact e-mail: \href{mailto:sarira@nucleares.unam.mx}{sarira@nucleares.unam.mx}},\ %
B.~Medina-Carrillo$^{2}$%
\thanks{Contact e-mail: \href{mailto:benjamin.medina@cinvestav.mx}{ benjamin.medina@cinvestav.mx}},\ %
G.~Sánchez-Colón$^{2}$%
\thanks{Contact e-mail: \href{mailto: gabriel.sanchez@cinvestav.mx}{gabriel.sanchez@cinvestav.mx}},\ %
Subhash Rajpoot$^{3}$%
\thanks{Contact e-mail: \href{mailto: Subhash.Rajpoot@csulb.edu}{Subhash.Rajpoot@csulb.edu}}%
\\
$^{1}$Instituto de Ciencias Nucleares, Universidad Nacional Aut\'onoma de M\'exico,\\ Circuito Exterior S/N, C.U., A.P. 70-543, CDMX 04510, México.\\
$^{2}$Departamento de Física Aplicada, Centro de Investigación y de Estudios Avanzados del IPN, Unidad Mérida.\\ A.P. 73, Cordemex, Mérida, Yucatán 97310, México.\\
$^{3}$Department of Physics and Astronomy, California State University,
1250 Bellflower Boulevard, Long Beach, CA 90840, USA.
}

\date{}

\pubyear{2023}

\begin{document}
\label{firstpage}
\pagerange{\pageref{firstpage}--\pageref{lastpage}}
\maketitle

\begin{abstract}
Observation of several very high energy (VHE) flaring events of the BL Lac object VER J0521+211 were reported by the VERITAS and MAGIC collaborations between 2009 and 2014. The redshift of this source is uncertain and several analysis have derived different limits for it. In the framework of the photohadronic model, and using three different extragalactic background light (EBL) models, we analyze seven independent VHE spectra of VER J0521+211 and determine the limiting values on its redshift. It is observed that the photohadronic scenario provides excellent fits to the reported observations. It is further observed that the photohadronic scenario, along with the EBL model of  Domínguez et al., puts the most restrictive limits on the redshift $z$ of VER J0521+211: $0.29\leq z \leq 0.31$ from the confidence level (CL) intervals at $2\sigma$, or a more conservative $0.28\leq z \leq 0.33$, at $3\sigma$. 
\end{abstract}



\newpage

\section{Introduction}
VER J0521+211 was first detected as a very high energy gamma-ray (VHE $> 100$ GeV) source by VERITAS in the year 2009~\citep{2009ATel.2260....1O}. The source is highly variable at all wavebands and is associated with the radio-loud active galaxy RGB J0521.8+2112. Several VHE flaring events from VER J0521+211 were observed between 2009 and 2014. During the 2009 VHE flaring period, the source exhibited a high-frequency BL Lac (HBL) like behavior with the synchrotron spectrum in the harder X-ray regime (in the range $10^{15}$ Hz to $10^{17}$ Hz)~\citep{Archambault_2013}. It was observed simultaneously in the multiwavelength during a sustained flaring episode from late 2012 to early 2014. This period also included the observations by MAGIC and VERITAS. For this entire observation period the synchrotron peak frequency remained between $10^{14}$ Hz and $10^{15}$ Hz and classified this source as an intermediate-frequency-peaked BL Lac (IBL) object. However, its redshift remained uncertain~\citep{VERITAS:2022htr}.

The VERITAS collaboration detected a VHE gamma-ray flare on 2020 February 25 from VER J0521+211 with steadily rising flux reaching a value $> 130\%$ of the VHE flux from the Crab Nebula~\citep{2020ATel13522....1Q}. This lead to the coordinated multiwavelength observations by several other instruments. During the flaring period, the Swift XRT observed~\citep{2020ATel13522....1Q} an order of magnitude increase in the X-ray flux compared to the one recorded weeks and months before. Similarly, the Fermi-LAT also observed elevated GeV gamma-rays~\citep{2020ATel13528....1S}. Contrary to this, the optical monitoring of VER J0521+211 showed a dimming of the source, suspecting an orphan flare which was not visible in optical wavelength~\citep{2020ATel13548....1P}. The IceCube neutrino telescope also looked for track-like muon neutrino events from the direction of VER J0521+211 but with null results~\citep{2020ATel13532....1P}.

In BL Lac objects the non thermal emission dominates over the stellar emission of the host galaxy and hinders the estimation of the redshift. The knowledge of the redshift of the source is essential to the understanding of the intrinsic spectrum, the nature of the source, and its cosmic evolution. Since VHE gamma-rays are attenuated by the EBL, knowledge on the redshift helps to probe the EBL effect through lepton pair production. The observed VHE flux of a blazar $F_{\gamma}$ is related to the intrinsic flux $F_{int}$, through the relation~\citep{Hauser:2001xs}
\beq
F_{\gamma}(E_{\gamma})=F_{int}(E_\gamma)\, e^{-\tau_{\gamma\gamma}(E_{\gamma}, z)},
\label{eq:flux}
\eeq
where $\tau_{\gamma\gamma}$ is the optical depth for the process $\gamma\gamma\rightarrow e^+e^-$ and depends on the photon energy $E_{\gamma}$ and the redshift $z$ of the source. The exponential factor corresponds to the depletion of the VHE flux due to the lepton pair-production. Thus, the distance to the source is essential to estimate the intrinsic flux from the observed flux. Well-known EBL models~\citep{Franceschini:2008tp,2010ApJ...712..238F,10.1111/j.1365-2966.2010.17631.x,10.1111/j.1365-2966.2012.20841.x} are used by the Imaging Atmospheric Cerenkov Telescopes (IACTs) collaborations to analyze the observed VHE $\gamma$-ray spectra from objects of varying redshifts.

The redshift of the IBL VER J0521+211 remains uncertain even after multiple measurements.~\cite{Shaw:2013pp} have reported the redshift of VER J0521+211 to be $z = 0.108$. However, subsequent studies do not confirm this value~\citep{Archambault_2013, Paiano:2017pol}. On the other hand,~\cite{Paiano:2017pol} have reported a lower limit of $z > 0.18$ and~\cite{Archambault_2013} have reported an upper limit of $z < 0.34$ from different analysis. Recently, by using several EBL models to construct the VHE spectra of the flaring events of 2013 and 2014, a $95\%$ confidence level upper limit of $z\leq 0.31$ was derived~\citep{VERITAS:2022htr}.

On the theoretical side, the photohadronic model of~\cite{Sahu:2019lwj} has been previously used to explain successfully the VHE spectra of 42 flaring epochs of 23 blazars; also, depending on the spectral index of the observed VHE $\gamma$-ray spectra, a classification scheme is proposed~\citep{Sahu_2019}. Using the same classification scheme of the VHE flaring events, VHE spectra of many HBLs and extreme HBLs (EHBLs) are also explained well. Additionally, due to the similarity between the blazar jet and the gamma-ray burst (GRB) jet, the photohadronic model is also well suited to explain the VHE spectra of GRB 190114C, GRB 180720B and GRB 190829A~\citep{Sahu:2020dsg, Sahu:2022qaw}. Recently, we have also used this model to interpret the VHE photons from GRB 221009A~\citep{2023ApJ...942L..30S}.

In this work, we use the photohadronic model of~\cite{Sahu:2019lwj} and the three well-known EBL models~\citep{Franceschini:2008tp,10.1111/j.1365-2966.2010.17631.x,10.1111/j.1365-2966.2012.20841.x} to analyze the VHE flaring events of 2009-2010 and 2013-2014 observed by VERITAS and MAGIC collaborations and obtain constraints on the limiting values of the redshift of the IBL VER J0521+211.

\section{Multiwavelength Observations of VER J0521+211}\label{sec2}

The VERITAS collaboration analyzed the previously undetected gamma-ray hot spots (above photon energy $> 30$ GeV) collected by Fermi-LAT in its first year of operations~\citep{Errando:2011ey}. The new VHE gamma-ray source, VER J0521+211,  was detected during the observation period 22 to 24 of October 2009 (MJD 55126-55128) and it is associated with RGB J0521.8+2112, a radio and X-ray source. VERITAS continued observing VER J0521+211 for a total exposure time of 14.5 hours between 2009 October 22 and 2010 January 16 and detected a high flux state on 2009 November 22 (MJD 55126-55212)~\citep{Archambault_2013}. VER J0521+211 is highly variable at all wavelengths with an average integral flux above 200 GeV and is among the brightest known TeV blazars. Afterwards, the broadband spectral energy distribution (SED) in the synchrotron range showed a  peak in the optical band which unambiguously classifies VER J0521+211 as an IBL. However, during the VHE flaring in November 2009~\citep{2009ATel.2260....1O}, VER J0521+211 also showed HBL-like properties.

During 2012 November to 2014 February, VER J0521+211 was observed in multiwavelength by VERITAS, MAGIC, Fermi-LAT, Swift-XRT and many other instruments and found to be in a prolonged gamma-ray flaring state~\citep{VERITAS:2022htr}. VERITAS observed the source with a total exposure of 23.6 hr (after data quality selection and dead time corrections). The Fermi-LAT observations to the source between 2012 October and 2014 May claimed to observe variability in the GeV gamma-ray light curve. However, during 2013 January 29 and 2014 January 24, Fermi-LAT observed a flux increase of about 11 times and the monthly flux was 7 times larger than the Fermi Large Area telescope Third Source Catalog (3FGL,~\citep{2015ApJS..218...23A}) value. This clearly demonstrates that the source was in a long-lasting elevated GeV gamma-ray flux state.

It was observed that the X-ray flux observed by the Swift-XRT (in the energy range 0.3 to 10 keV) and the TeV gamma-ray flux had a moderate correlation. The X-ray spectrum appeared harder when the flux was higher. Also, it is important to note that there was no significant increase in the optical flux during the observation period as compared to the archival low state flux.

MAGIC observations on VER J0521+211 were undertaken on four nights, October 15 and 16, November 29, and December 2 of 2013 with a total effective time of $\sim 4.5$ hr. On the night of December 3, VERITAS observed a peak flux above 200 GeV with no intraday variability. However, the night before, i.e., on December 2, MAGIC observed a lower flux above 200 GeV~\citep{VERITAS:2022htr}.

~\cite{VERITAS:2022htr} reported a Bayesian block analysis (a modeling technique that finds the optimal segmentation of the data in the observation interval~\citep{Scargle_2013}) of the VHE spectral curves observed by VERITAS and MAGIC. Two change points (the edges of the Bayesian blocks) were obtained on 2013 December 2 and 2013 December 6, which define three Bayesian blocks as BB1 (MJD 56580.0 to MJD 56628.5; 2013 December 2 to  2013 December 6), BB2 (MJD 56628.5 to MJD 56632.5; 2013 October 15 to 2013 December 2), and BB3 (MJD 56632.5 to MJD 56689.0; 2013 December 6 to 2014 February 1). The flux in each block is constant with no flux variability. The BB2 interval has only one night observation on 2013 December 3 by VERITAS for 2.3 hr and its average flux is $\sim 37\%$ of the Crab Nebula flux, the highest among all the Bayesian blocks. Taking into account the Bayesian block analysis for the three time intervals, the multiwavelength SEDs constructed using four different EBL models, a conservative 95\% confidence upper limit on the redshift, $z\le 0.31$, was found by~\cite{VERITAS:2022htr}.
    
\section{\label{sec3}The Photohadronic model}

HBLs are important sources of VHE gamma-rays and flaring in this energy regime seems to be the major activity of many HBLs. The flaring periods also vary widely, ranging from few minutes to weeks, and at the same time switching from quiescent state to very active state and vice versa~\citep{Albert:2007zd,Aharonian:2007ig,Ghisellini:2008gn,Ghisellini:2008us,2009A&A...502..499R,Senturk:2013pa}. The VHE emission mechanisms are still not understood properly~\citep{Krawczynski:2003fq,VERITAS:2004avc,Blazejowski:2005ih,Aharonian_2009,HESS:2011huh} and different models such as leptonic, hadronic, and hybrid, have been developed to explain these flaring events~\citep{Giannios:2009pi, Cerruti:2014iwa}. Most of these models have a large number of free parameters which limit their predictability power. Those with less number of free parameters usually face other difficulties such as having large magnetic fields and/or invoking super-Eddington luminosity. The broadband SEDs are constructed from many simultaneous multiwavelength observations to constrain different theoretical models~\citep{Senturk:2013pa, MAGIC:2016tfe}. Furthermore, the propagating VHE gamma-rays suffer energy-depending attenuation on their way to Earth by the intervening EBL through $e^{\pm}$ production~\citep{Stecker:1992wi, Fermi-LAT:2012qrj, Padovani:2017zpf}. As a result, the shape of the observed VHE spectrum is modified.

The salient feature of the photohadronic scenario is that the flaring is explained by assuming the formation of a double jet structure, one jet enclosing the other jet along the same axis~\citep{Sahu:2019lwj, Sahu_2019}. The inner jet of size $R'_f$ and a photon density of $n'_{\gamma,f}$ is buried inside the outer jet of size $R'_b$ ($R'_b > R'_f$) having a photon density $n'_{\gamma}$, with $n'_{\gamma} \ll n'_{\gamma,f}$ (the prime implies the co-moving frame). We assume that the internal and the external jets are moving with almost the same bulk Lorentz factor $\Gamma$ and with a common Doppler factor ${\cal D}$. The geometrical structure of the jet is depicted in Figure 1 of~\cite{Sahu:2019lwj}. In a photohadronic scenario, the Fermi accelerated high energy protons are injected into the inner jet region with a power-law spectrum $dN/dE_{p} \propto E^{-\alpha}_{p}$, where $E_{p}$ is the proton energy and the proton spectral index $\alpha \ge 2$~\citep{1993ApJ...416..458D}. These protons interact with the  synchrotron self-Compton (SSC) background seed photons of the inner jet and produce a $\Delta$-resonance which subsequently decays to gamma-rays and neutrinos via intermediate $\pi^0$ and $\pi^+$ production, respectively. As the inner jet region is hidden and there is no direct way to estimate the photon density in the inner jet region, we assume a scaling behavior between the inner and the outer jet regions, which can be expressed as~\citep{Sahu:2015tua}
\beq
\frac{n'_{\gamma,f}(\epsilon_{\gamma,1})}{n'_{\gamma,f}(\epsilon_{\gamma,2})} \simeq\frac{n'_{\gamma}(\epsilon_{\gamma,1})}{n'_{\gamma}(\epsilon_{\gamma,2})}.
\label{eq:scaling}
\eeq
The above equation shows that the ratio of the photon densities at two different background energies $\epsilon_{\gamma,1}$ and $\epsilon_{\gamma,2}$ in the flare region and the non-flaring region are almost the same. However, the ratio of the photon densities in the non-flaring region (right hand side of Eq.~(\ref{eq:scaling})) can be calculated from the observed SED. Thus, the ratio of the photon densities in the inner jet region can be inferred. Finally, the observed VHE photon flux is given by
\beq
F_{\gamma}(E_{\gamma})=F_0 \left ( \frac{E_\gamma}{\rm TeV} \right )^{-\delta+3}\,e^{-\tau_{\gamma\gamma}(E_\gamma,z)}.
\label{eq:fluxgeneral}
\eeq
By comparing Eqs.~(\ref{eq:flux}) and~(\ref{eq:fluxgeneral}), the intrinsic flux is deduced to be 
\beq
F_{int}(E_{\gamma})=F_0 \left ( \frac{E_\gamma}{\rm TeV} \right )^{-\delta+3}.
\label{eq:fluxintrinsic}
\eeq 
Here, the spectral index of the VHE photon $\delta=\alpha+\beta$ (with $\beta$ the seed photon spectral index) is the only free parameter in the model, the normalization factor $F_0$ can be calculated from the observed data. The value of $\delta$ should be in the range $2.5\le \delta \le 3.0$~\citep{Sahu_2019}.

\section{\label{sec4}Analysis and Results}

The VERITAS and the MAGIC collaborations observed primarily seven VHE gamma-rays flaring epochs from the IBL VER J0521+211 between 2009 and 2014. The source has an unknown redshift and several analysis have been done to constraint the redshift. We use the photohadronic model along with three well-known EBL models to fit these seven VHE spectra. In Eq.~(\ref{eq:fluxgeneral}), the normalization constant $F_0$, the spectral index $\delta$, and the redshift $z$, are considered as free parameters and are simultaneously varied to find their best fit values. For all the three EBL models considered, the estimated parameters for the best fits are presented in Table~\ref{table1} (from third to fifth column). Using the best fit values of $z$, $F_0$, and $\delta$ from the three EBL models, the redshift CL intervals at $1\sigma$, $2\sigma$, and $3\sigma$ are determined and displayed in columns sixth, seventh, and eight of Table~\ref{table1}, respectively. As stated before, for our analysis we constrain the spectral index $\delta$ in the range $2.5\le \delta \le 3.0$. For convenience, we have defined the VERITAS observations between MJD 55126 (2009 October 22) and MJD 55212 (2010 January 16)~\citep{Archambault_2013} as V1. Also, we have define the MAGIC VHE spectra measured on three nights, 2013 October 15, October 16, and November 29 as MO15, MO16, and MN29, respectively. The three Bayesian blocks BB1, BB2, and BB3 are described in Sec.~\ref{sec2}. Again, for convenience and simplicity, we define EBL models as consisting of the models of Franceschini~(\cite{Franceschini:2008tp}), Gilmore~(\cite{10.1111/j.1365-2966.2012.20841.x}) and Domínguez~(\cite{10.1111/j.1365-2966.2010.17631.x}).


\begin{figure*}
\includegraphics[width=7in]{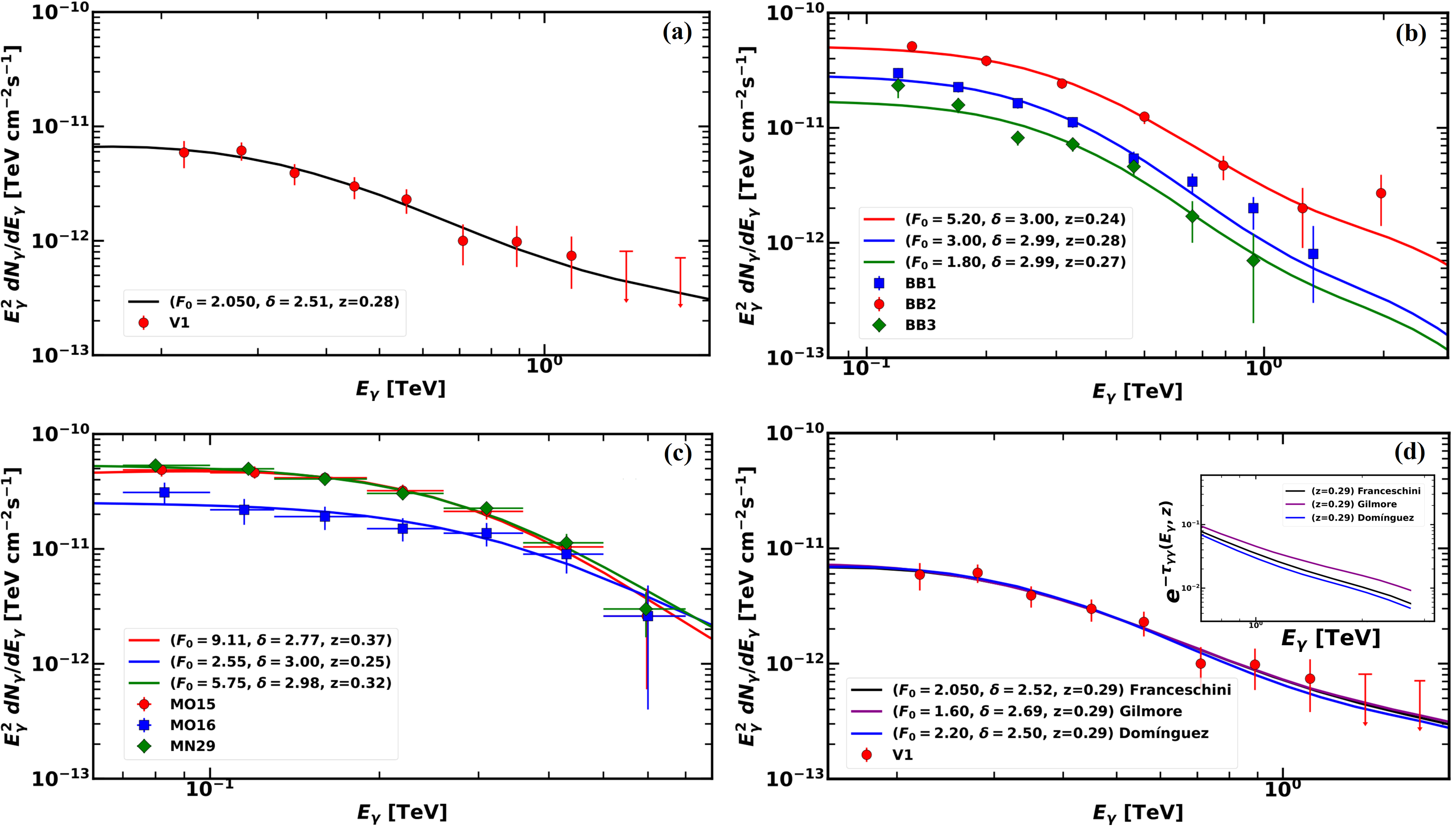}
\caption{Fits of all the seven observed VHE $\gamma$-ray spectra using the photohadronic model along with the EBL model of Domínguez are given as follows (a) The time averaged spectrum V1, (b) The observed VHE $\gamma$-ray spectra BB1, BB2, and BB3, and (c) The VHE $\gamma$-ray spectra MO15, MO16, and MN29. (d) The time averaged spectrum V1 is fitted for $z=0.29$ using the three EBL models for comparison.
The region for $E_{\gamma} \gtrsim 0.7$ TeV is zoomed in to demonstrate the differences between the different EBL model predictions of $e^{-\tau_{\gamma\gamma}}$. This is shown in the inset. For all cases, values of $F_0$ (in units of $10^{-11} \mathrm{TeV\, cm^{-2}\, s^{-1}}$), $\delta$, and $z$ are given in the sub-figures.}
\label{fig:Fig1}
\end{figure*}

According to the classification scheme given in~\cite{Sahu_2019}, the VHE emission states of a HBL are defined as the very high emission state for $2.5 \leq \delta \leq 2.6$, the high emission state for $2.6 < \delta < 3.0$, and the low emission state when $\delta=3.0$. It can be seen that the time-averaged spectrum V1 (given in Table~\ref{table1}) is in the very high emission state as fitted by all the EBL models. However, the EBL model of Gilmore predicts slightly higher value of $\delta=2.62$. The redshift $z$ predicted by each EBL model for the flaring events  seems to differ slightly from each other. 

We use the best fit parameters given in Table~\ref{table1} for the EBL model of~\cite{10.1111/j.1365-2966.2010.17631.x} to fit the observed spectra which are shown in Fig.~\ref{fig:Fig1} (a), (b), and (c). In Fig.~\ref{fig:Fig1} (a) the fit to the VHE spectrum V1 is shown with $z=0.28$, $F_0=2.05\times 10^{-11}\,\mathrm{TeV\, cm^{-2}\, s^{-1}}$, and $\delta=2.51$; this corresponds to a very high emission state. It is to be noted that between 2009 October 22 and 2010 January 16 for about three months period, the flaring was in a very high emission state. This might also be a favorable time window to look for high energy neutrinos from VER J0521+11 direction in the offline data of IceCube 59 string configuration (IC59)~\citep{2013ApJ...779..132A,2014PhRvD..89f2007A}. In Fig.~\ref{fig:Fig1} (b) we have fitted the VHE spectra BB1, BB2 and BB3 and their corresponding ($\delta$, $z$) values are $(2.99, 0.28)$, $(3.0, 0.24)$, and $(2.99, 0.27)$ respectively. These values of $\delta$ show that the spectra are mostly in the low emission states. In Fig.~\ref{fig:Fig1} (c) the VHE spectra MO15, MO16, and MN29 are fitted with their respective best fit parameters as given in Table~\ref{table1}. 

To compare the predictions of the three EBL models considered, we fixed $z=0.29$ and fitted the VHE spectrum V1 with the photohadronic model along with these EBL models which are shown in Fig.~\ref{fig:Fig1} (d). All the three EBL models fit very well to the data. However, by zooming in for $E_{\gamma} \gtrsim 0.7$ TeV (the inset in Fig.~\ref{fig:Fig1} (d)), we have shown the VHE photon survival factor $e^{-\tau_{\gamma\gamma}}$ for all the three models. The EBL model of~\cite{10.1111/j.1365-2966.2012.20841.x} gives slightly higher value. For a complete analysis, the best fits to the VHE spectra of different observation periods by VERITAS and MAGIC given in Table~\ref{table1} for the EBL models of~\cite{Franceschini:2008tp} and~\cite{10.1111/j.1365-2966.2012.20841.x} are presented in the Appendix in the figures A1 to A6.


\begin{table*}
\centering
\caption{Estimated values of the parameters of the photohadronic model obtained from the best fits to the VHE spectra of VER J0521+211 are given. We refer to the main text for the different observations given in the first column and also for different EBL models shown in the second column. $F_0$ (in units of $ 10^{-11}\,\mathrm{TeV\, cm^{-2}\, s^{-1}}$), the spectral index $\delta$, and the redshift $z$ are given in the third, the fourth, and the fifth columns, respectively. Columns six, seven and eight give the range for the redshift values at one, two and three sigma confidence levels as reported in the text.
}
\label{table1}
\begin{tabular}{cccccccc}
\hline
VER J0521+211 & EBL & \multicolumn{3}{c}{Parameters} & \multicolumn{3}{c}{Redshift CL intervals} \\
Observations & Model & $F_0$ & $\delta$ & $z$ & $1\sigma$ & $2\sigma$ & $3\sigma$\\
\hline
V1 & Franceschini & 2.05 & 2.52 & 0.29 & (0.20, 0.32) & (0.18, 0.35) & (0.17, 0.36)\\
 & Gilmore & 1.80 & 2.62 & 0.30 & (0.22, 0.35) & (0.19, 0.38) & (0.18, 0.39)\\
 & Domínguez & 2.05 & 2.51 & 0.28 & (0.19, 0.31) & (0.17, 0.33) & (0.16, 0.34)\\
BB1 & Franceschini & 2.90 & 3.00 & 0.28 & (0.27, 0.31) & (0.26, 0.34) & (0.25, 0.36)\\
 & Gilmore & 3.10 & 3.00 & 0.31 & (0.29, 0.33) & (0.28, 0.37) & (0.27, 0.39)\\
 & Domínguez & 3.00 & 2.99 & 0.28 & (0.26, 0.30) & (0.25, 0.33) & (0.24, 0.34)\\
BB2 & Franceschini & 5.30 & 3.00 & 0.25 & (0.24, 0.29) & (0.23, 0.33) & (0.22, 0.35)\\
 & Gilmore & 5.45 & 3.00 & 0.27 & (0.26, 0.31) & (0.25, 0.36) & (0.24, 0.38)\\
 & Domínguez & 5.20 & 3.00 & 0.24 & (0.23, 0.27) & (0.22, 0.31) & (0.21, 0.33)\\
BB3 & Franceschini & 1.80 & 3.00 & 0.28 & (0.25, 0.31) & (0.24, 0.35) & (0.23, 0.37)\\
 & Gilmore & 1.85 & 3.00 & 0.30 & (0.27, 0.34) & (0.26, 0.38) & (0.25, 0.41)\\
 & Domínguez & 1.80 & 2.99 & 0.27 & (0.25, 0.30) & (0.23, 0.33) & (0.22, 0.35)\\
MO15 & Franceschini & 8.50 & 2.79 & 0.37 & (0.31, 0.44) & (0.29, 0.47) & (0.28, 0.47)\\
 & Gilmore & 11.1 & 2.70 & 0.42 & (0.34, 0.48) & (0.31, 0.51) & (0.30, 0.52)\\
 & Domínguez & 9.10 & 2.77 & 0.37 & (0.30, 0.43) & (0.28, 0.47) & (0.27, 0.48)\\
MO16 & Franceschini & 2.53 & 3.00 & 0.26 & (0.21, 0.33) & (0.18, 0.40) & (0.17, 0.45)\\
 & Gilmore & 2.65 & 2.99 & 0.28 & (0.23, 0.36) & (0.19, 0.44) & (0.18, 0.48)\\
 & Domínguez & 2.55 & 3.00 & 0.25 & (0.21, 0.32) & (0.18, 0.39) & (0.17, 0.43)\\
MN29 & Franceschini & 6.55 & 2.93 & 0.34 & (0.31, 0.39) & (0.30, 0.43) & (0.29, 0.44)\\
 & Gilmore & 5.90 & 2.99 & 0.35 & (0.33, 0.42) & (0.29, 0.47) & (0.31, 0.48)\\
 & Domínguez & 5.75 & 2.98 & 0.32 & (0.30, 0.37) & (0.29, 0.41) & (0.28, 0.43)\\
\hline
\end{tabular}
\end{table*}

\begin{table*}
\centering
\caption{The redshift overlapping regions of all the seven observations shown in Table~\ref{table1} for $2\sigma$ CL intervals and for $3\sigma$ CL intervals are calculated separately for each EBL model. There is no overlapping region for $1\sigma$ CL intervals.}
\label{table2}
\begin{tabular}{lccc}
\hline
EBL & $1\sigma$ & $2\sigma$ & $3\sigma$\\
\hline
Franceschini & - & (0.30, 0.33) & (0.29, 0.35)\\
Gilmore & - & (0.32, 0.36) & (0.31, 0.38)\\ 
Domínguez & - & (0.29, 0.31) & (0.28, 0.33)\\
\hline
\end{tabular}
\end{table*}

\begin{figure*}
\centering
\includegraphics[width=4in]{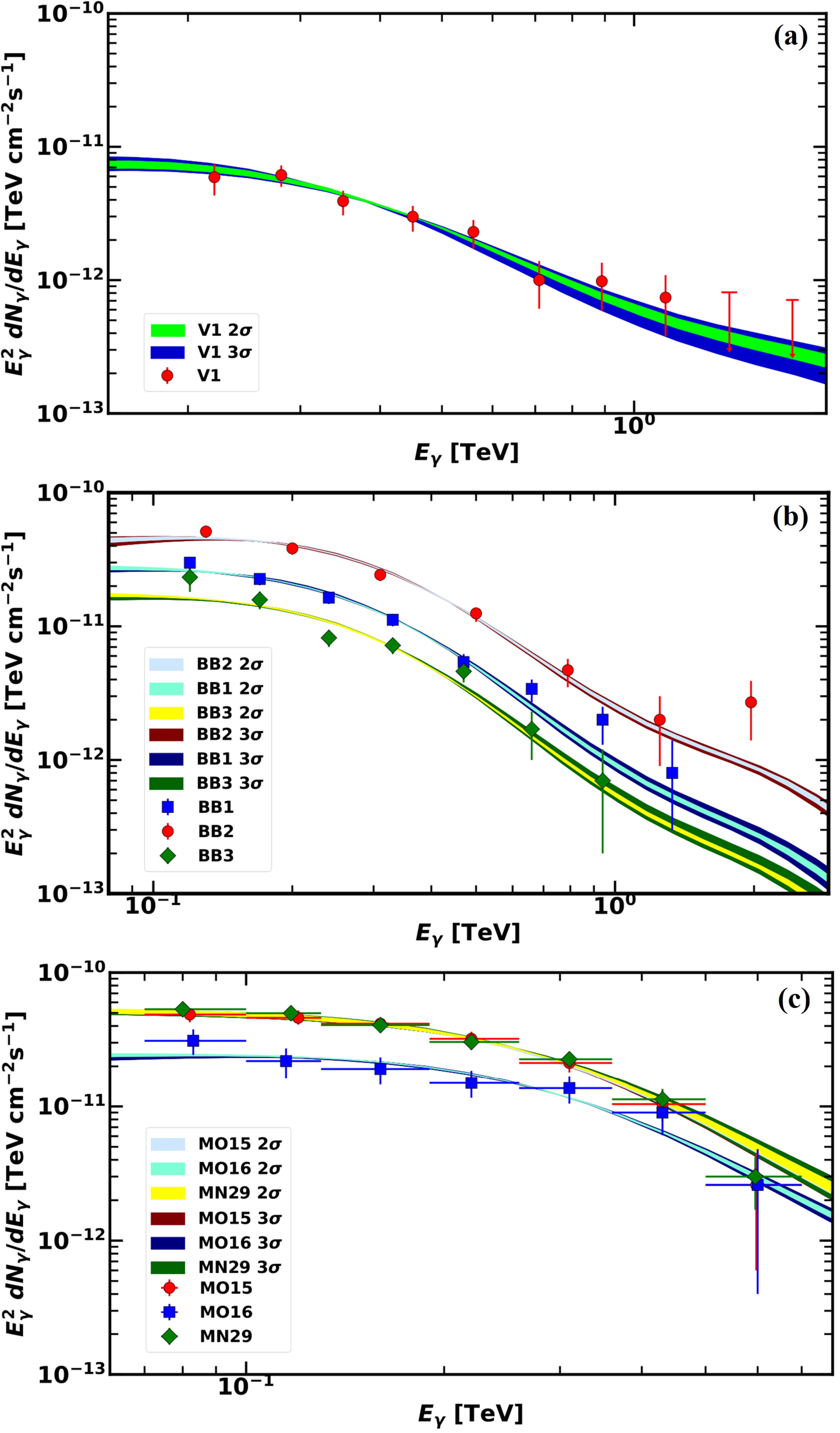}
\caption{All the seven observed VHE spectra of VER J0521+211 are fitted separately using the photohadronic model at the limiting values of the overlapping regions of $z$ for $2\sigma$ ($0.29, \, 0.31$) and for $3\sigma$ ($0.28, \, 0.33$) CL intervals using the EBL model of~Domínguez. 
In (a) V1 spectrum is given; BB1, BB2, and BB3 are in (b) and the spectra MO15, MO16 and MN29 are in (c).}
\label{fig:Fig2}
\end{figure*}

For a given EBL model we superimpose the $2\sigma$ CL intervals of $z$ for all the seven observations given in Table~\ref{table1} and find the overlapping region from them. We repeat this procedure for $1\sigma$ and $3\sigma$ CL intervals for all EBL models separately. The overlapping region gives the limiting values of the redshift corresponding to that particular $\sigma$ CL interval and the EBL model. The lower bound is the highest lower value of $z$ and the upper bound is the lowest upper value of $z$ respectively. However, we do not find these limits for $z$ in the $1\sigma$ CL intervals. For the three EBL models, the overlapping regions of the redshift CL intervals are determined and displayed in Table~\ref{table2}. The analysis shows that the EBL model of~\cite{10.1111/j.1365-2966.2010.17631.x} provides the most stringent constraint on $z$. At $2\sigma$ and at $3\sigma$ CL, the constraints are $0.29 \le z \le 0.31$ and $0.28 \le z \le 0.33$, respectively. So far, no other analysis has given such a stringent lower limit of $z$ for the IBL VER J0521+211.

The photohadronic model uses the limiting values of the redshift from the EBL model of~\cite{10.1111/j.1365-2966.2010.17631.x} given in  Table~\ref{table2} to fit the observed spectra as shown in Fig.~\ref{fig:Fig2}. In  Fig.~\ref{fig:Fig2} (a) V1 spectrum is given; BB1, BB2, and BB3 are in Fig.~\ref{fig:Fig2} (b) and the spectra MO15, MO16 and MN29 are in Fig.~\ref{fig:Fig2} (c). All these plots clearly show that the photohadronic fits are very good when we consider $2\sigma$ and/or $3\sigma$ overlapping regions.

Our analysis shows that the flaring events of the IBL VER J0521+211 can be fitted very well for $2.5\le \delta \le 3.0$ even though the classification scheme and the range of $\delta$ are applicable for the VHE spectra of a HBL~\citep{Sahu_2019}. Also, it is important to note that our analysis is based on the fitting to the VHE spectra of seven observations.

\section{\label{sec5}Discussions}

Between 2009 and 2014, VERITAS and MAGIC collaborations observed seven VHE $\gamma$-ray flaring epochs from the IBL VER J0521+211 whose redshift is yet to be confirmed. However, using the multiwavelength observations, several analysis were undertaken to constraint the redshift of this source. The photohadronic model~\citep{Sahu:2019lwj,Sahu_2019} has successfully explained the VHE $\gamma$-ray flaring events from several HBLs and EHBLs. Previously, this model has also been used to constraint the redshift of several HBLs with unknown $z$~\citep{Sahu_2019}. For the first time the same photohadronic model is used to analyze the VHE spectra of an IBL. Here, we employ the photohadronic model along with three well-known EBL models for a detailed analysis of the VHE spectra of the source VER J0521+211 and to impose stringent constraint on its redshift. We show that we can fit the observed spectra well by using the three EBL models for different redshifts. The uncertainties in the redshift determination that are dependent on the photohadronic model parameters ($F_0$ and $\delta$) are displayed in the last three columns of Table 1, at $1\sigma$, $2\sigma$, and $3\sigma$ individual CL intervals. This has been done for each one of the seven independent VHE observations of VER J0521+511 and for each EBL model considered. However, the CL intervals to $z$ at different sigmas decide which EBL model gives the most stringent limits on $z$. Our analysis shows that the photohadronic model with EBL corrections from the EBL model of~\cite{10.1111/j.1365-2966.2010.17631.x} gives the overlapping regions $0.29 \le z \le 0.31$ at $2\sigma$ CL intervals and $0.28\le z \le 0.33$ at $3\sigma$ CL intervals. To our understanding, the lower limit of $0.29$ at $2\sigma$ and $0.28$ at $3\sigma$ CL intervals are the most stringent lower limits to $z$ obtained so far for the IBL VER J0521+211. It is to be emphasized that our conclusion is based on the fitting to the VHE spectra of seven observations of VER J0521+211. Our expectations are that the photohadronic model will also be able to fit the VHE spectra of many IBLs using the classification scheme adopted for HBLs. 

As noted previously, the IBL VER J0521+211 was in a very high emission state for about three months period during 2009 October 22 and 2010 January 16. Three months are sufficient to accelerate protons to very high energies and to produce high energy neutrinos by  their interactions with the ambient photons. Thus, we propose to look for high energy neutrinos from VER J0521+11 direction in the offline data of IC59.

Additionally, the coordinated multiwavelength observations after the detection of VHE gamma-ray flare on 2020 February 25, conveys the message that VER J0521+211 is an important IBL which needs to be studied further in detail to understand its emission mechanisms and can also be a potential source of high energy neutrinos that should be looked for.

\section*{Acknowledgements}
The work of S.S. is partially supported by DGAPA-UNAM (México) Project No. IN103522. B. M-C and G. S-C would like to thank CONACyT (México) for partial support. Partial support from CSU-Long Beach is gratefully acknowledged. 


\section*{Data Availability}
No new data were generated or analysed in support of this research.


\bibliographystyle{mnras}
\bibliography{ref}

\begin{thebibliography}{}
\makeatletter
\relax
\def\mn@urlcharsother{\let\do\@makeother \do\$\do\&\do\#\do\^\do\_\do\%\do\~}
\def\mn@doi{\begingroup\mn@urlcharsother \@ifnextchar [ {\mn@doi@}
  {\mn@doi@[]}}
\def\mn@doi@[#1]#2{\def\@tempa{#1}\ifx\@tempa\@empty \href
  {http://dx.doi.org/#2} {doi:#2}\else \href {http://dx.doi.org/#2} {#1}\fi
  \endgroup}
\def\mn@eprint#1#2{\mn@eprint@#1:#2::\@nil}
\def\mn@eprint@arXiv#1{\href {http://arxiv.org/abs/#1} {{\tt arXiv:#1}}}
\def\mn@eprint@dblp#1{\href {http://dblp.uni-trier.de/rec/bibtex/#1.xml}
  {dblp:#1}}
\def\mn@eprint@#1:#2:#3:#4\@nil{\def\@tempa {#1}\def\@tempb {#2}\def\@tempc
  {#3}\ifx \@tempc \@empty \let \@tempc \@tempb \let \@tempb \@tempa \fi \ifx
  \@tempb \@empty \def\@tempb {arXiv}\fi \@ifundefined
  {mn@eprint@\@tempb}{\@tempb:\@tempc}{\expandafter \expandafter \csname
  mn@eprint@\@tempb\endcsname \expandafter{\@tempc}}}

\bibitem[\protect\citeauthoryear{{Aartsen} et~al.,}{{Aartsen}
  et~al.}{2013}]{2013ApJ...779..132A}
{Aartsen} M.~G.,  et~al., 2013, \mn@doi [\apj] {10.1088/0004-637X/779/2/132},
  \href {https://ui.adsabs.harvard.edu/abs/2013ApJ...779..132A} {779, 132}

\bibitem[\protect\citeauthoryear{{Aartsen} et~al.,}{{Aartsen}
  et~al.}{2014}]{2014PhRvD..89f2007A}
{Aartsen} M.~G.,  et~al., 2014, \mn@doi [\prd] {10.1103/PhysRevD.89.062007},
  \href {https://ui.adsabs.harvard.edu/abs/2014PhRvD..89f2007A} {89, 062007}

\bibitem[\protect\citeauthoryear{Abramowski et~al.}{Abramowski
  et~al.}{2012}]{HESS:2011huh}
Abramowski A.,  et~al., 2012, \mn@doi [ApJ.] {10.1088/0004-637X/746/2/151},
  746, 151

\bibitem[\protect\citeauthoryear{{Acero} et~al.,}{{Acero}
  et~al.}{2015}]{2015ApJS..218...23A}
{Acero} F.,  et~al., 2015, \mn@doi [ApJS] {10.1088/0067-0049/218/2/23}, \href
  {https://ui.adsabs.harvard.edu/abs/2015ApJS..218...23A} {218, 23}

\bibitem[\protect\citeauthoryear{Ackermann et~al.}{Ackermann
  et~al.}{2012}]{Fermi-LAT:2012qrj}
Ackermann M.,  et~al., 2012, \mn@doi [Science] {10.1126/science.1227160}, 338,
  1190

\bibitem[\protect\citeauthoryear{Adams et~al.}{Adams
  et~al.}{2022}]{VERITAS:2022htr}
Adams C.~B.,  et~al., 2022, \mn@doi [ApJ.] {10.3847/1538-4357/ac6dd9}, 932, 129

\bibitem[\protect\citeauthoryear{Aharonian et~al.}{Aharonian
  et~al.}{2007}]{Aharonian:2007ig}
Aharonian F.,  et~al., 2007, \mn@doi [ApJ.] {10.1086/520635}, 664, L71

\bibitem[\protect\citeauthoryear{Aharonian et~al.,}{Aharonian
  et~al.}{2009}]{Aharonian_2009}
Aharonian F.,  et~al., 2009, \mn@doi [ApJ.] {10.1088/0004-637X/695/1/L40}, 695,
  L40

\bibitem[\protect\citeauthoryear{Ahnen et~al.}{Ahnen
  et~al.}{2017}]{MAGIC:2016tfe}
Ahnen M.~L.,  et~al., 2017, \mn@doi [A\&A.] {10.1051/0004-6361/201629960}, 603,
  A29

\bibitem[\protect\citeauthoryear{Albert et~al.}{Albert
  et~al.}{2007}]{Albert:2007zd}
Albert J.,  et~al., 2007, \mn@doi [ApJ.] {10.1086/521382}, 669, 862

\bibitem[\protect\citeauthoryear{Archambault et~al.,}{Archambault
  et~al.}{2013}]{Archambault_2013}
Archambault S.,  et~al., 2013, \mn@doi [ApJ.] {10.1088/0004-637x/776/2/69},
  776, 69

\bibitem[\protect\citeauthoryear{Blazejowski et~al.}{Blazejowski
  et~al.}{2005}]{Blazejowski:2005ih}
Blazejowski M.,  et~al., 2005, \mn@doi [ApJ.] {10.1086/431925}, 630, 130

\bibitem[\protect\citeauthoryear{Cerruti, Zech, Boisson  \& Inoue}{Cerruti
  et~al.}{2015}]{Cerruti:2014iwa}
Cerruti M.,  Zech A.,  Boisson C.,   Inoue S.,  2015, \mn@doi [MNRAS.]
  {10.1093/mnras/stu2691}, 448, 910

\bibitem[\protect\citeauthoryear{Cui et~al.,}{Cui
  et~al.}{2005}]{VERITAS:2004avc}
Cui W.,  et~al., 2005, \mn@doi [AIP Conf. Proc.] {10.1063/1.1878445}, 745, 455

\bibitem[\protect\citeauthoryear{{Dermer} \& {Schlickeiser}}{{Dermer} \&
  {Schlickeiser}}{1993}]{1993ApJ...416..458D}
{Dermer} C.~D.,  {Schlickeiser} R.,  1993, \mn@doi [ApJ.] {10.1086/173251},
  \href {https://ui.adsabs.harvard.edu/abs/1993ApJ...416..458D} {416, 458}

\bibitem[\protect\citeauthoryear{Domínguez et~al.,}{Domínguez
  et~al.}{2011}]{10.1111/j.1365-2966.2010.17631.x}
Domínguez A.,  et~al., 2011, \mn@doi [MNRAS.]
  {10.1111/j.1365-2966.2010.17631.x}, 410, 2556

\bibitem[\protect\citeauthoryear{Errando}{Errando}{2011}]{Errando:2011ey}
Errando M.,  2011, in {3rd International Fermi Symposium}.  (\mn@eprint {arXiv}
  {1111.1209})

\bibitem[\protect\citeauthoryear{{Finke}, {Razzaque}  \& {Dermer}}{{Finke}
  et~al.}{2010}]{2010ApJ...712..238F}
{Finke} J.~D.,  {Razzaque} S.,   {Dermer} C.~D.,  2010, \mn@doi [ApJ.]
  {10.1088/0004-637X/712/1/238}, \href
  {https://ui.adsabs.harvard.edu/abs/2010ApJ...712..238F} {712, 238}

\bibitem[\protect\citeauthoryear{Franceschini, Rodighiero  \&
  Vaccari}{Franceschini et~al.}{2008}]{Franceschini:2008tp}
Franceschini A.,  Rodighiero G.,   Vaccari M.,  2008, \mn@doi [A\&A.]
  {10.1051/0004-6361:200809691}, 487, 837

\bibitem[\protect\citeauthoryear{Ghisellini \& Tavecchio}{Ghisellini \&
  Tavecchio}{2008}]{Ghisellini:2008gn}
Ghisellini G.,  Tavecchio F.,  2008, \mn@doi [MNRAS.]
  {10.1111/j.1745-3933.2008.00454.x}, 386, 28

\bibitem[\protect\citeauthoryear{Ghisellini, Tavecchio, Bodo  \&
  Celotti}{Ghisellini et~al.}{2009}]{Ghisellini:2008us}
Ghisellini G.,  Tavecchio F.,  Bodo G.,   Celotti A.,  2009, \mn@doi [MNRAS.]
  {10.1111/j.1745-3933.2008.00589.x}, 393, 16

\bibitem[\protect\citeauthoryear{Giannios, Uzdensky  \& Begelman}{Giannios
  et~al.}{2010}]{Giannios:2009pi}
Giannios D.,  Uzdensky D.~A.,   Begelman M.~C.,  2010, \mn@doi [MNRAS.]
  {10.1111/j.1365-2966.2009.16045.x}, 402, 1649

\bibitem[\protect\citeauthoryear{Gilmore, Somerville, Primack  \&
  Domínguez}{Gilmore et~al.}{2012}]{10.1111/j.1365-2966.2012.20841.x}
Gilmore R.~C.,  Somerville R.~S.,  Primack J.~R.,   Domínguez A.,  2012,
  \mn@doi [MNRAS.] {10.1111/j.1365-2966.2012.20841.x}, 422, 3189

\bibitem[\protect\citeauthoryear{Hauser \& Dwek}{Hauser \&
  Dwek}{2001}]{Hauser:2001xs}
Hauser M.~G.,  Dwek E.,  2001, \mn@doi [ARA\&A.]
  {10.1146/annurev.astro.39.1.249}, 39, 249

\bibitem[\protect\citeauthoryear{Krawczynski et~al.}{Krawczynski
  et~al.}{2004}]{Krawczynski:2003fq}
Krawczynski H.,  et~al., 2004, \mn@doi [ApJ.] {10.1086/380393}, 601, 151

\bibitem[\protect\citeauthoryear{{Ong}}{{Ong}}{2009}]{2009ATel.2260....1O}
{Ong} R.~A.,  2009, ATel., \href
  {https://ui.adsabs.harvard.edu/abs/2009ATel.2260....1O} {2260, 1}

\bibitem[\protect\citeauthoryear{Padovani et~al.}{Padovani
  et~al.}{2017}]{Padovani:2017zpf}
Padovani P.,  et~al., 2017, \mn@doi [A\&AR.] {10.1007/s00159-017-0102-9}, 25, 2

\bibitem[\protect\citeauthoryear{Paiano, Landoni, Falomo, Treves, Scarpa  \&
  Righi}{Paiano et~al.}{2017}]{Paiano:2017pol}
Paiano S.,  Landoni M.,  Falomo R.,  Treves A.,  Scarpa R.,   Righi C.,  2017,
  \mn@doi [ApJ.] {10.3847/1538-4357/837/2/144}, 837, 144

\bibitem[\protect\citeauthoryear{{Pizzuto} \& {Taboada}}{{Pizzuto} \&
  {Taboada}}{2020}]{2020ATel13532....1P}
{Pizzuto} A.,  {Taboada} I.,  2020, ATel., \href
  {https://ui.adsabs.harvard.edu/abs/2020ATel13532....1P} {13532, 1}

\bibitem[\protect\citeauthoryear{{Pursimo} \& {Ojha}}{{Pursimo} \&
  {Ojha}}{2020}]{2020ATel13548....1P}
{Pursimo} T.,  {Ojha} R.,  2020, ATel., \href
  {https://ui.adsabs.harvard.edu/abs/2020ATel13548....1P} {13548, 1}

\bibitem[\protect\citeauthoryear{{Quinn} \& {VERITAS Collaboration}}{{Quinn} \&
  {VERITAS Collaboration}}{2020}]{2020ATel13522....1Q}
{Quinn} J.,  {VERITAS Collaboration} 2020, ATel., \href
  {https://ui.adsabs.harvard.edu/abs/2020ATel13522....1Q} {13522, 1}

\bibitem[\protect\citeauthoryear{{Resconi}, {Franco}, {Gross}, {Costamante}  \&
  {Flaccomio}}{{Resconi} et~al.}{2009}]{2009A&A...502..499R}
{Resconi} E.,  {Franco} D.,  {Gross} A.,  {Costamante} L.,   {Flaccomio} E.,
  2009, \mn@doi [A\&A.] {10.1051/0004-6361/200911770}, \href
  {https://ui.adsabs.harvard.edu/abs/2009A&A...502..499R} {502, 499}

\bibitem[\protect\citeauthoryear{Sahu}{Sahu}{2019}]{Sahu:2019lwj}
Sahu S.,  2019, \mn@doi [Rev. Mex. Fis.] {10.31349/revmexfis.65.307}, 65, 307

\bibitem[\protect\citeauthoryear{Sahu \& L\'opez~Fort\'\i{}n}{Sahu \&
  L\'opez~Fort\'\i{}n}{2020}]{Sahu:2020dsg}
Sahu S.,  L\'opez~Fort\'\i{}n C.~E.,  2020, \mn@doi [ApJL.]
  {10.3847/2041-8213/ab93da}, 895, L41

\bibitem[\protect\citeauthoryear{Sahu, Miranda  \& Rajpoot}{Sahu
  et~al.}{2016}]{Sahu:2015tua}
Sahu S.,  Miranda L.~S.,   Rajpoot S.,  2016, \mn@doi [EPJ. C]
  {10.1140/epjc/s10052-016-3975-2}, 76, 127

\bibitem[\protect\citeauthoryear{Sahu, Fort{\'{\i}}n  \& Nagataki}{Sahu
  et~al.}{2019}]{Sahu_2019}
Sahu S.,  Fort{\'{\i}}n C. E.~L.,   Nagataki S.,  2019, \mn@doi [ApJ.]
  {10.3847/2041-8213/ab43c7}, 884, L17

\bibitem[\protect\citeauthoryear{Sahu, Polanco  \& Rajpoot}{Sahu
  et~al.}{2022}]{Sahu:2022qaw}
Sahu S.,  Polanco I. A.~V.,   Rajpoot S.,  2022, \mn@doi [ApJ.]
  {10.3847/1538-4357/ac5cc6}, 929, 70

\bibitem[\protect\citeauthoryear{{Sahu}, {Medina-Carrillo},
  {S{\'a}nchez-Col{\'o}n}  \& {Rajpoot}}{{Sahu}
  et~al.}{2023}]{2023ApJ...942L..30S}
{Sahu} S.,  {Medina-Carrillo} B.,  {S{\'a}nchez-Col{\'o}n} G.,   {Rajpoot} S.,
  2023, \mn@doi [\apjl] {10.3847/2041-8213/acac2f}, \href
  {https://ui.adsabs.harvard.edu/abs/2023ApJ...942L..30S} {942, L30}

\bibitem[\protect\citeauthoryear{Scargle, Norris, Jackson  \& Chiang}{Scargle
  et~al.}{2013}]{Scargle_2013}
Scargle J.~D.,  Norris J.~P.,  Jackson B.,   Chiang J.,  2013, \mn@doi [ApJ.]
  {10.1088/0004-637x/764/2/167}, 764, 167

\bibitem[\protect\citeauthoryear{Sent\"urk, Errando, B\"ottcher  \&
  Mukherjee}{Sent\"urk et~al.}{2013}]{Senturk:2013pa}
Sent\"urk G.~D.,  Errando M.,  B\"ottcher M.,   Mukherjee R.,  2013, \mn@doi
  [ApJ.] {10.1088/0004-637X/764/2/119}, 764, 119

\bibitem[\protect\citeauthoryear{Shaw et~al.,}{Shaw et~al.}{2013}]{Shaw:2013pp}
Shaw M.~S.,  et~al., 2013, \mn@doi [ApJ.] {10.1088/0004-637X/764/2/135}, 764,
  135

\bibitem[\protect\citeauthoryear{{Sinapius}, {Angioni}  \& {Ojha}}{{Sinapius}
  et~al.}{2020}]{2020ATel13528....1S}
{Sinapius} J.,  {Angioni} R.,   {Ojha} R.,  2020, ATel., \href
  {https://ui.adsabs.harvard.edu/abs/2020ATel13528....1S} {13528, 1}

\bibitem[\protect\citeauthoryear{Stecker, de Jager  \& Salamon}{Stecker
  et~al.}{1992}]{Stecker:1992wi}
Stecker F.~W.,  de Jager O.~C.,   Salamon M.~H.,  1992, \mn@doi [ApJL.]
  {10.1086/186369}, 390, L49

\makeatother
\end{thebibliography}


\newpage
\appendix

\section{Franceschini and Gilmore EBL models.}

In this appendix, the plots for the best fits to the VHE spectra of VER J0521+211 with the photohadronic model and the EBL models of~\cite{Franceschini:2008tp} and~\cite{10.1111/j.1365-2966.2012.20841.x}, are presented. In all fits presented, $F_0$ is in units of $10^{-11} \mathrm{TeV\, cm^{-2}\, s^{-1}}$.

\begin{figure*}
\includegraphics[width=4.5in]{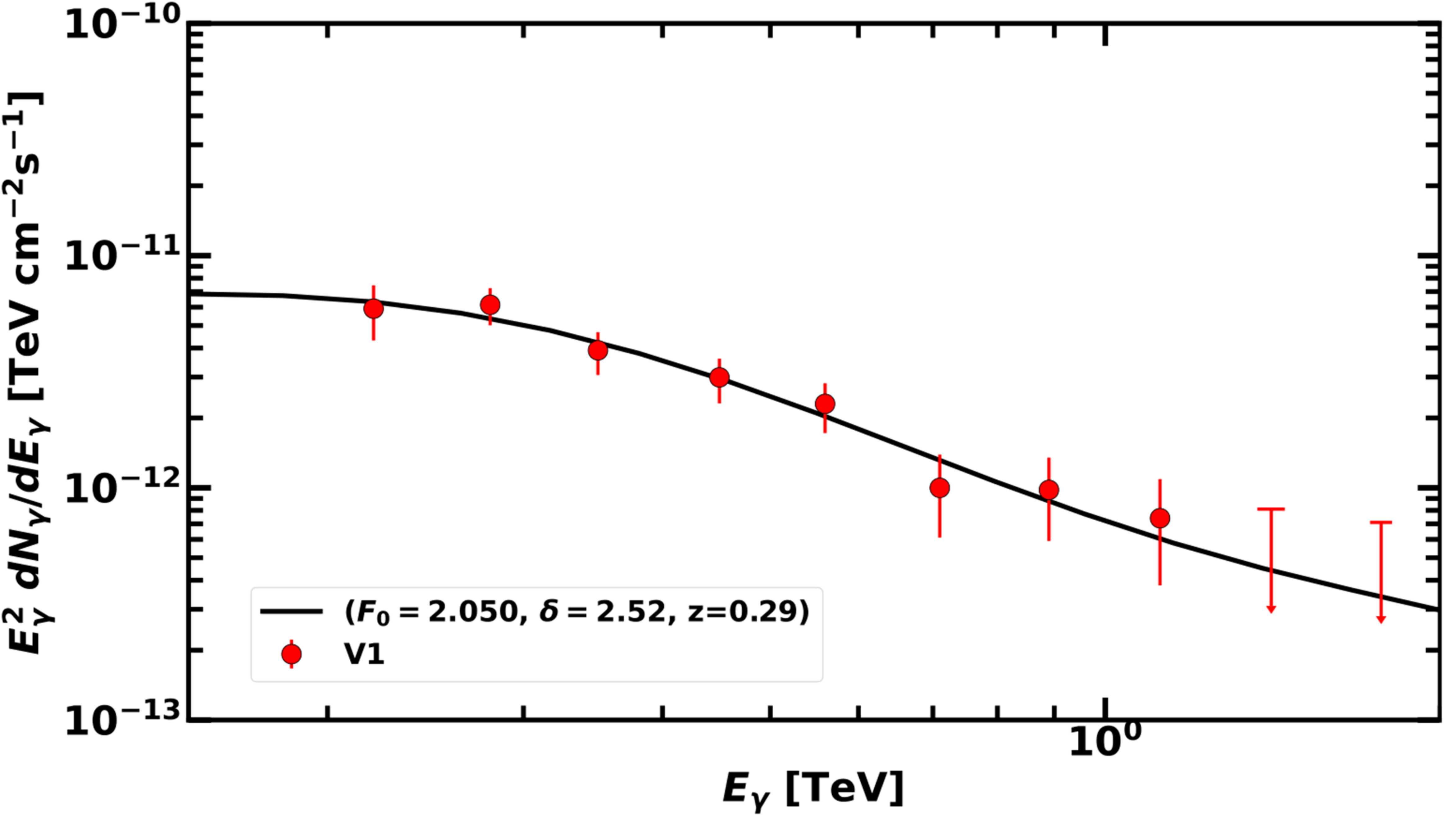}
\caption{The time-averaged VHE spectrum V1 (defined in the main text) is fitted with the photohadronic model by incorporating the EBL model of~Franceschini.}
\end{figure*}

\begin{figure*}
\includegraphics[width=4.5in]{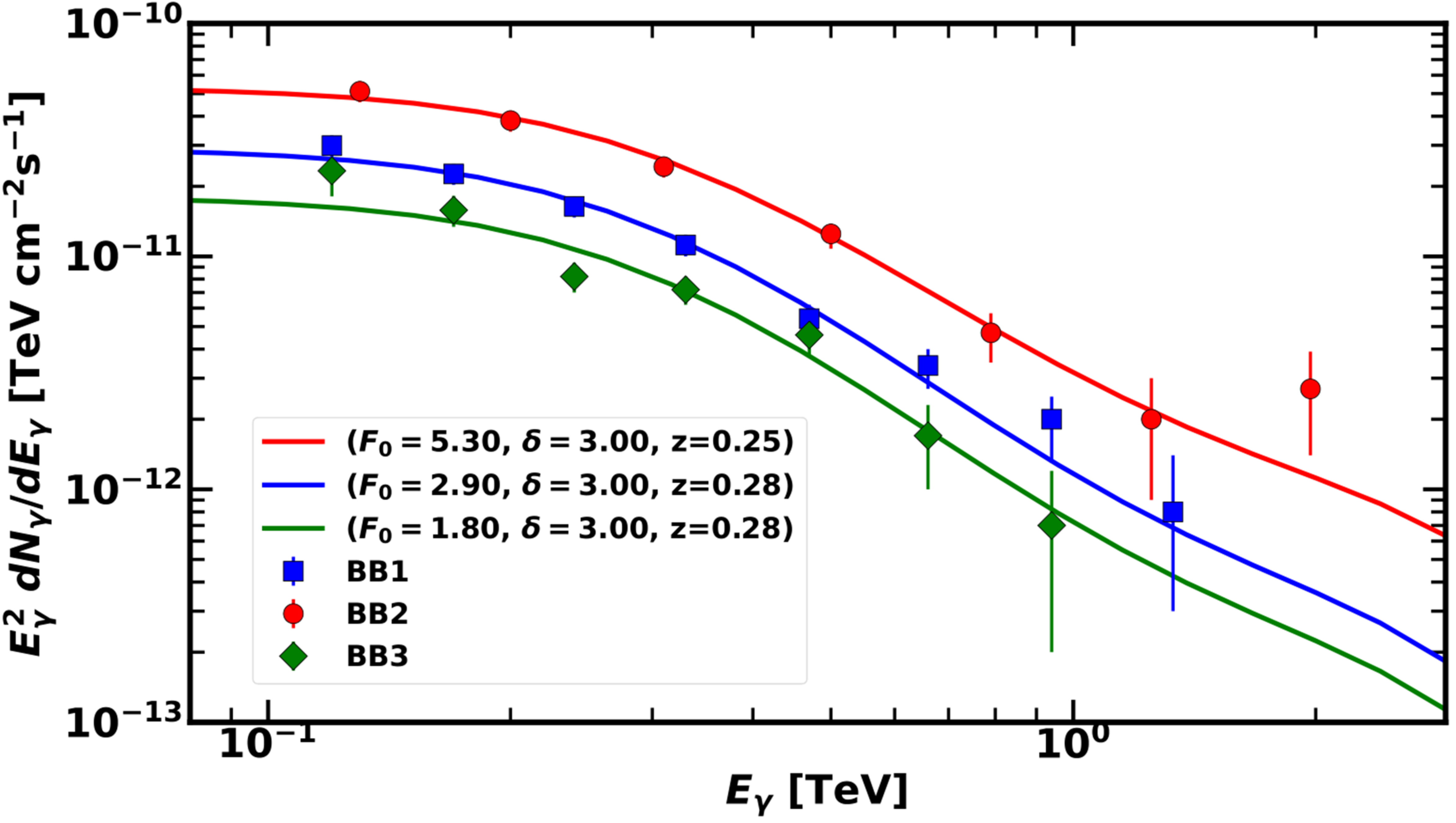}
\caption{The photohadronic fit including the EBL model of~Franceschini to the VHE spectra BB1, BB2 and BB3~\citep{VERITAS:2022htr}.
}
\end{figure*}

\begin{figure*}
\includegraphics[width=4.5in]{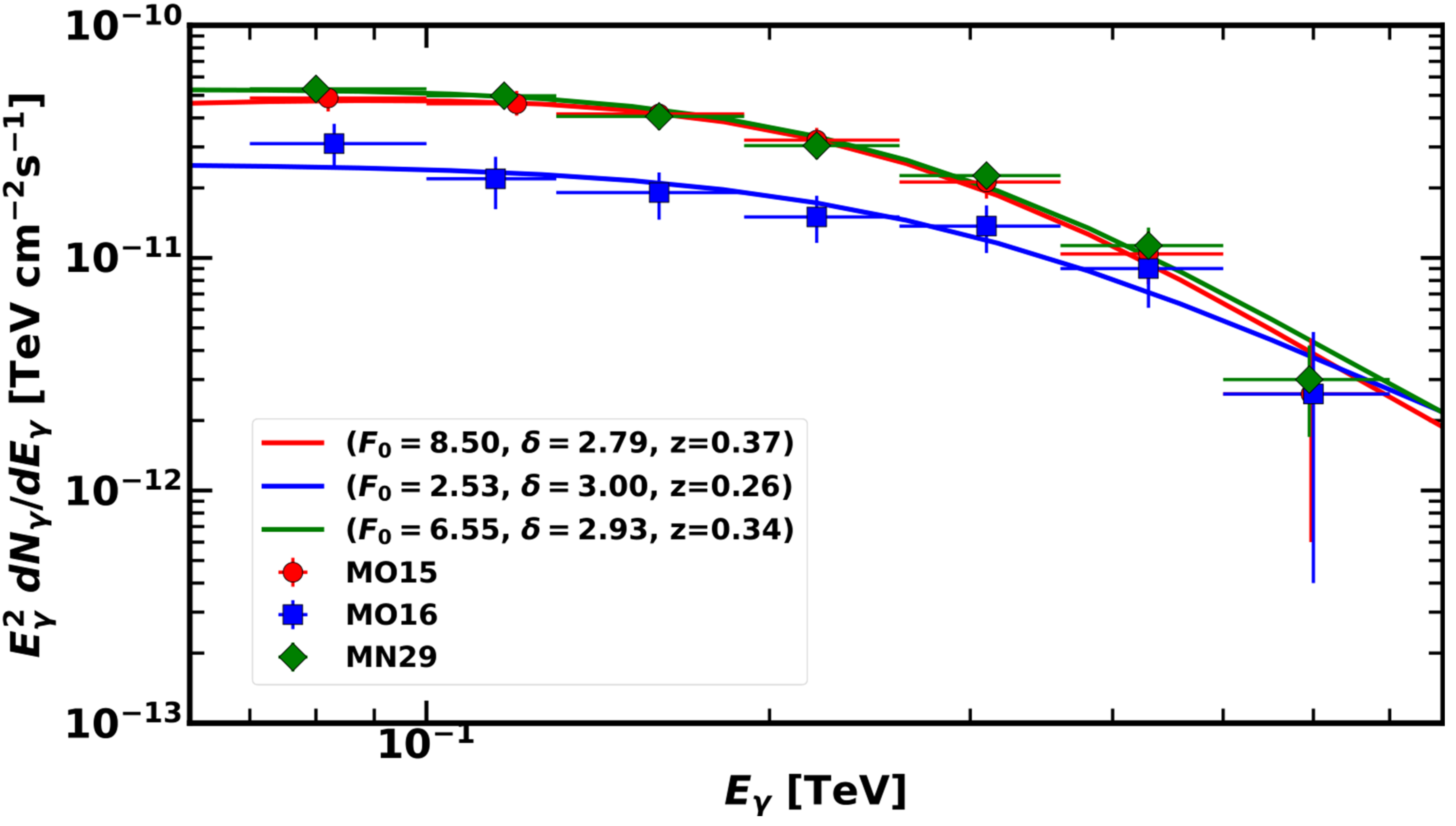}
\caption{The photohadronic fit including the EBL model of~Franceschini to the VHE spectra MO15, MO16 and MN29~\citep{VERITAS:2022htr}.
}
\end{figure*}

\begin{figure*}
\includegraphics[width=4.5in]{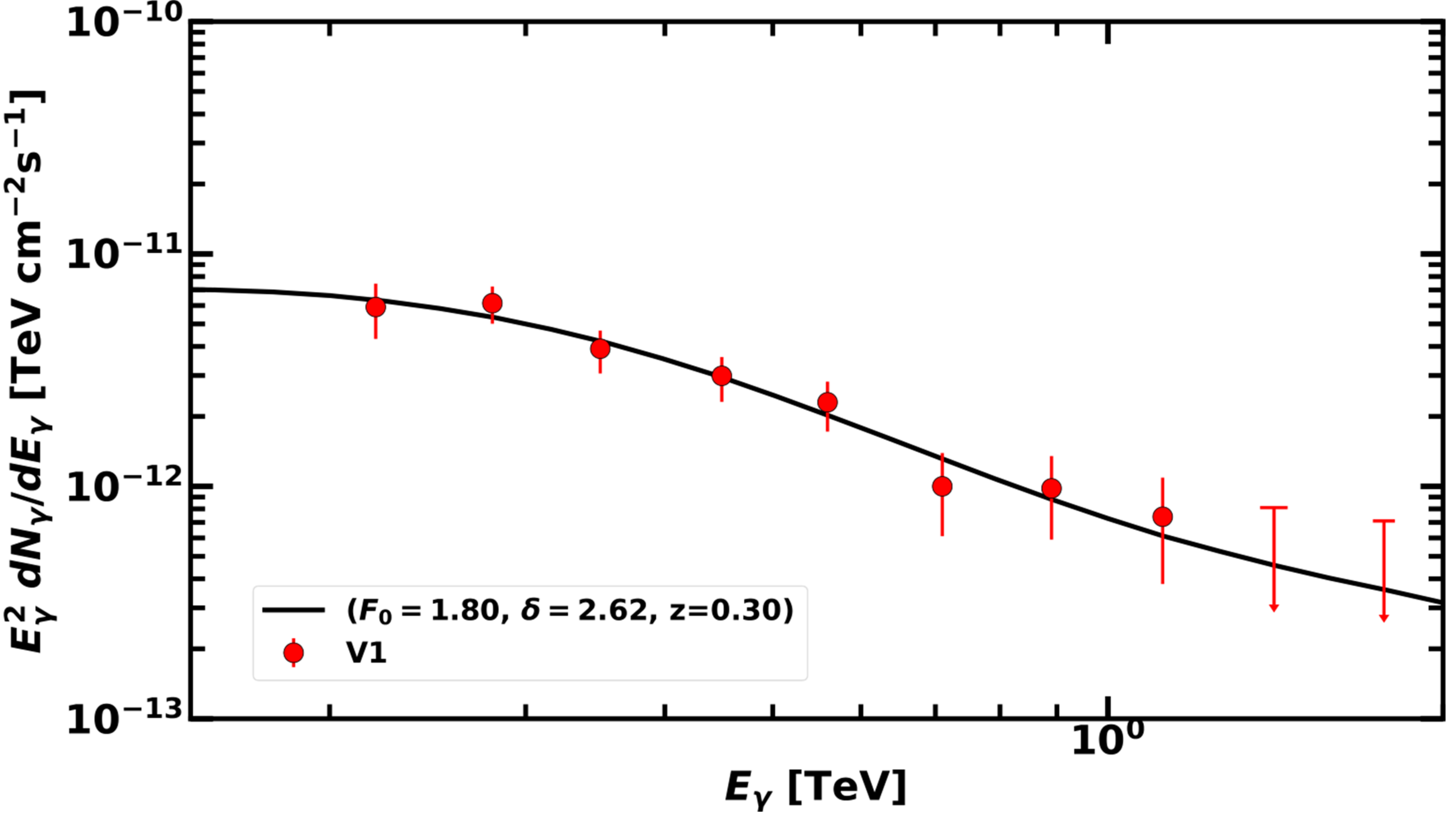}
\caption{Fitting to the VHE spectra V1 using the photohadronic model + EBL model of~Gilmore.}
\end{figure*}

\begin{figure*}
\includegraphics[width=4.5in]{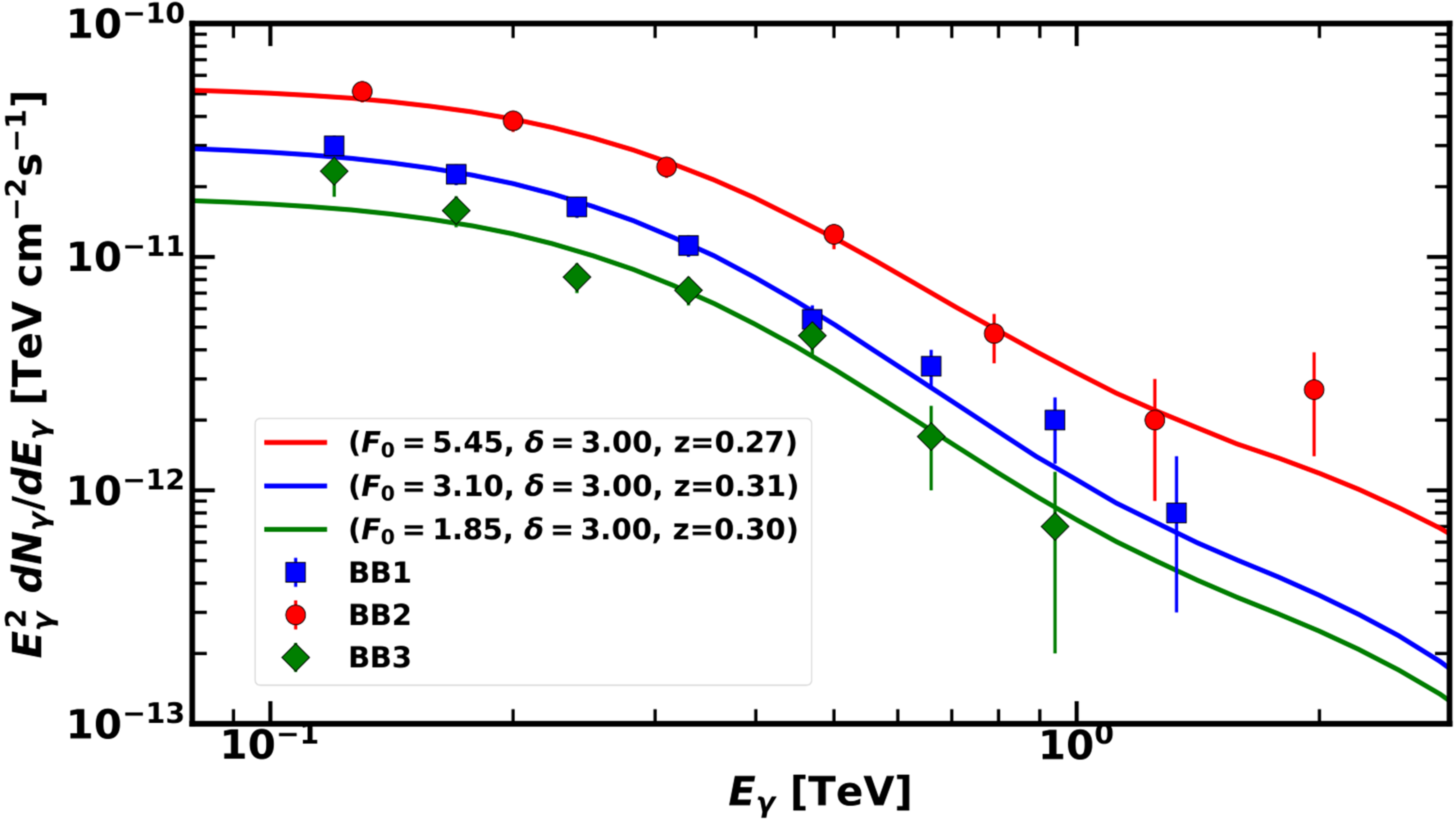}
\caption{The VHE spectra BB1, BB2 and BB3 are fitted by including the EBL model of~Gilmore to the photohadronic model.
}
\end{figure*}

\begin{figure*}
\includegraphics[width=4.5in]{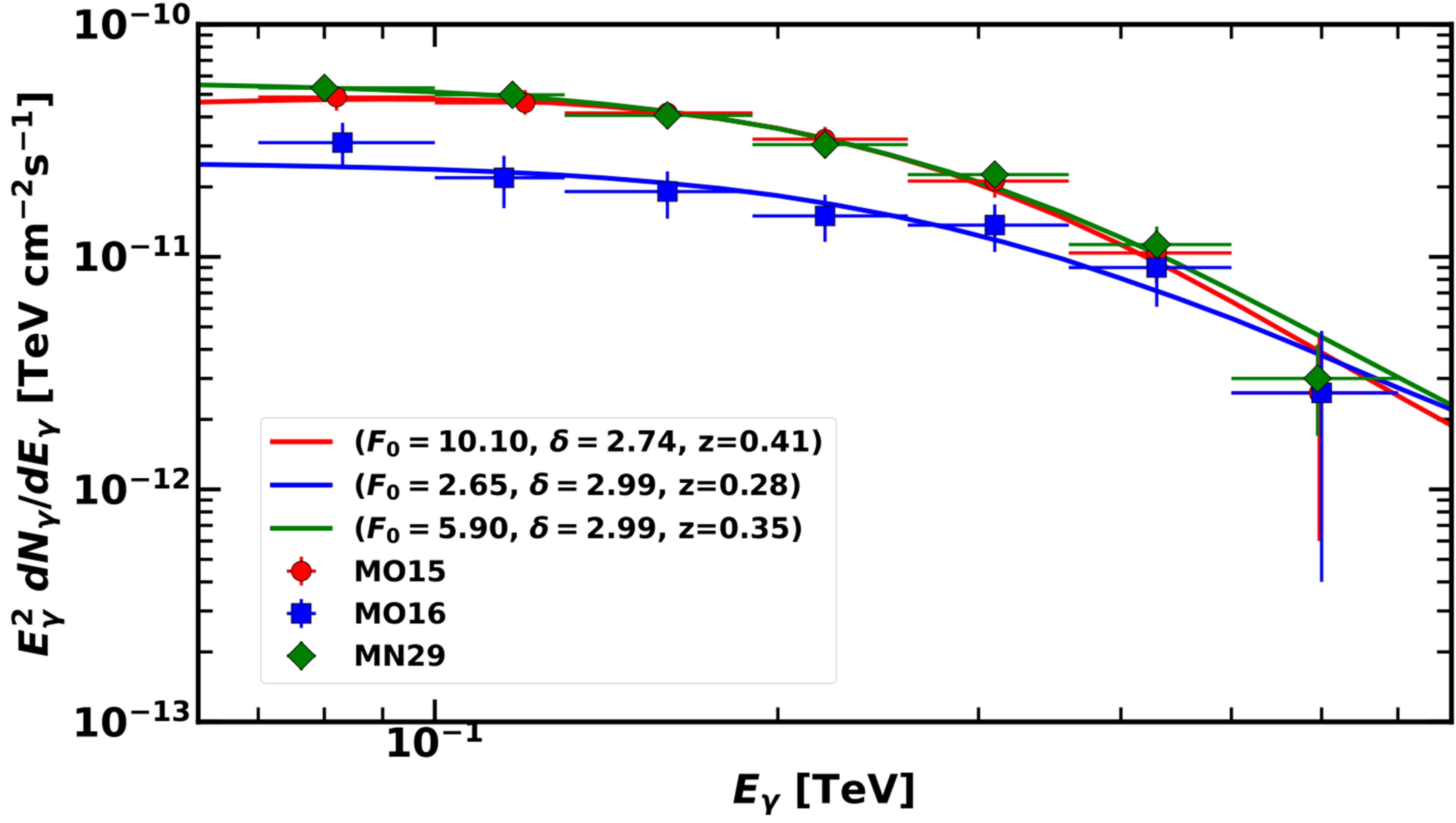}
\caption{The VHE spectra MO15, MO16 and MN29 are fitted by including the EBL model of~Gilmore to the photohadronic model.
}
\end{figure*}

\bsp	
\label{lastpage}
\end{document}